\documentclass[a4paper,11pt]{article}
\pdfoutput=1 

\usepackage{jcappub} 
\usepackage{multirow}                 

\usepackage{float}
\usepackage{subfig}
\usepackage[T1]{fontenc} 
\usepackage[table,xcdraw]{xcolor}

\title{Strange magnetars admixed with fermionic dark matter}

 \author{Osvaldo Ferreira$^{a}$}
 \author{and Eduardo S. Fraga$^{a}$}

\affiliation{$^{a}$Instituto de F\'\i sica, Universidade Federal do Rio de Janeiro,\\ Av. Athos da Silveira Ramos, 149, CEP 21941-909, Rio de Janeiro, Brazil}

\emailAdd{osvaldofn@pos.if.ufrj.br}
\emailAdd{fraga@if.ufrj.br }

\abstract{We discuss strange stars admixed with fermionic dark matter in the presence of a strong magnetic field using the two-fluid Tolman-Oppenheimer-Volkov equations. We describe strange quark matter using the MIT bag model and consider magnetic fields in the range $\sim 10^{17}-10^{18}$ G. For the fermionic dark matter, we consider the cases of free particles and strongly self-interacting particles, with dark fermion masses $m=5, 100, 500$ GeV. We discuss the effects of dark matter and a strong magnetic field on the masses and radii of the stars, as well as on its tidal deformability. Even though strong magnetic fields contribute to decreasing the total mass of the star, they attenuate the rate of decrease in the maximum mass brought about by increasing the dark matter fraction in the admixed system. The most intensely affected observable, however, is the tidal deformability, with variations on the range of $70\%-90\%$ for reasonable values of the magnetic field or dark matter central energy density.}

\begin{document}
\maketitle
\flushbottom

\section{Introduction} \label{sec:intro}

Compact stars exhibit the most extreme conditions in baryonic density that nuclear matter can endure \cite{Schaffner-Bielich:2020psc,Fukushima:2010bq,Graeff2019}. This renders their description challenging and fascinating. Some of these objects, magnetars, can also be under the influence of magnetic fields that can reach the range $B \sim 10^{17}-10^{18}$ G, among the highest realized in Nature \cite{Duncan:1992hi,Thompson:1993hn,Kouveliotou:1998ze}. Moreover, being so dense, compact stars are natural candidates to accrete dark matter (DM) \cite{Kouvaris:2010jy, dm_compact-constraints,CS_DM_probes,Constr_ADM_CS,Implications_ADM_CS,SQM_DM, DM_NS_to_SS, DCO_overview,Mukhopadhyay:2015xhs,Jimenez:2021nmr,Bell2019,Bell2020,Acevedo2020}. The presence of DM will naturally affect the structure of such admixed compact stars. On the other hand, observables related to those stars can, in principle, provide constraints on DM candidates.

Asymptotic freedom makes the presence of quark matter in the core of neutron stars very likely if central energy densities are high enough \cite{Annala:2019puf}. Therefore, one naturally expects to find hybrid stars and, perhaps, quark stars \cite{Ivanenko:1965dg,Itoh:1970uw} as stable branches in a mass-radius diagram. Following a seminal work by Witten \cite{Witten:1984rs}, a rich phenomenology of self-bound strange stars \cite{Alcock:1986hz,Haensel:1986qb} and quark (hybrid) stars naturally emerged using the MIT bag model \cite{Farhi:1984qu} as a framework for the EoS at high densities. Given its relevance for the physics of high dense matter, strange quark stars have been extensively studied and are still a highly active topic (examples of recent investigations can be found in Refs. \cite{SS1, SS2, SS3, SS4, SS5, SS6, SS7, SS8, SS9}). For a review on quark matter in neutron stars, see Ref. \cite{Buballa:2014jta}.

In this paper we consider strange stars admixed with fermionic dark matter in the presence of a strong magnetic field using the two-fluid Tolman-Oppenheimer-Volkov equations. We describe strange quark matter using the MIT bag model\footnote{We choose the framework of the MIT bag model motivated only by the possibility of direct comparison to previous work. Also, for the analysis that depends on a range of values for the dark fermion mass, magnetic field and the intensity of DM self-interaction, it is  convenient to avoid other degrees of freedom in parameters that would be necessary in more realistic descriptions of the EoS, such as those relying on perturbative QCD \cite{Fraga:2001id,Fraga:2004gz,Kurkela:2009gj,Fraga:2013qra,Kurkela:2014vha,Fraga:2015xha,Ghisoiu:2016swa,Annala:2017llu,Gorda:2018gpy,Annala:2019puf,Gorda:2021kme}. Even though results on perturbative magnetic QCD are available for very large magnetic fields \cite{Blaizot:2012sd,Sedaghat:2022fue}, we postpone this analysis to a future publication.}, 
and consider magnetic fields in the range $\sim 10^{17}-10^{18}$ G. For the fermionic dark matter, we consider the cases of free particles and strongly self-interacting particles, with dark fermion masses $m=5, 100, 500$ GeV. 

Neutron stars admixed with DM have been previously studied (see Ref. \cite{DelPopolo:2020hel} for a more complete list of references). Reference \cite{Tolos:2015qra} considers hybrid NS with an EoS for neutron star matter that uses perturbative QCD and effective field theory as high and low-density descriptions, respectively, and polytropes as interpolating functions, as discussed in Ref. \cite{Kurkela:2014vha}, besides taking into account inner and outer crusts and limits on the DM content of the stars in order to satisfy the two-solar mass observational constraint \cite{Demorest:2010bx,Antoniadis:2013pzd}. Reference \cite{Deliyergiyev:2019vti} extends these results to a wider range of dark fermion masses, from $1$ GeV to $500$ GeV. 

Strange stars admixed with dark matter have also been considered previously \cite{Mukhopadhyay:2015xhs,Jimenez:2021nmr}. However, the role of a strong magnetic field, as commonly found in pulsars,  has not been investigated so far in such admixed stars. In most cases these magnetic effects can indeed be neglected when computing stellar structural properties. However, in more extreme cases, as in magnetars, the very strong magnetic fields may play an important role \cite{intensity_mag1, effect2, effect3}. There, surface magnetic fields of intensities of order $10^{14}$-$10^{15}$G are required to explain astronomical observations \cite{Duncan:1992hi,Thompson:1993hn, Review_Mag1, Review_Mag2}. Moreover, the magnetic fields in the core of such stars may reach even higher values, of order $10^{17}-10^{18}$ G \cite{intensity_mag1,intensity_mag2}. 

Magnetars exhibit such extreme magnetic fields that several interesting phenomena are associated with them: they can probe strong field quantum electrodynamics effects \cite{Gorbar2021,Heyl2018}, axion-like-particles \cite{Fortin2021}, dense matter physics \cite{Negreiros2018}, etc. Effects of high magnetic fields on strange stars have been discussed previously in \cite{chak1,chak2,artigo_mag_deformabilidade,Deb2021, SSMag1, SSMag2, SSMag3}.

This paper is organized as follows. In section \ref{sec:structure_eq} we present and briefly discuss the structure equations for stars admixed with dark matter, the two-fluid Tolman-Oppenheimer-Volkov equations. In section \ref{sec: EoSs} we discuss the equations of state for magnetized strange quark matter and for dark matter. Section \ref{sec:results} contains our main results and discussion. Section \ref{sec:conclusion} presents our summary and perspectives. We adopt natural units, i.e. $\hbar=c=1$.

\section{Structure Equations for Compact Stars Admixed with Dark Matter}\label{sec:structure_eq}



We assume compact stars admixed with dark matter to be spherically symmetric and static systems. The matter and DM components are considered as two ideal fluids that interact only gravitationally. We also assume that the energy-momentum tensor is conserved for each fluid, independently. Therefore, both fluids contribute to the gravitational potential felt by the matter in the star, but the pressure gradient of a given component does not exert a direct force on the other component. The resulting structure equations are the so called two-fluid Tolman–Oppenheimer–Volkoff (TOV) equations \cite{Two_fluid_TOV, two_fluid_tov2,Ivanytskyi2020}:
\begin{equation}\label{2F_TOV1}
   \begin{aligned}
\frac{d p_{1}}{d r}=&-\frac{G M(r) \epsilon_{1}(r)}{r^{2}}\left[1+\frac{p_{1}(r)}{\epsilon_{1}(r)}\right]\left[1+4 \pi r^{3} \frac{\left(p_{1}(r)+p_{2}(r)\right)}{M(r)}\right]\left[1-2 G \frac{M(r)}{r}\right]^{-1} ,
\end{aligned} 
\end{equation}
\begin{equation}\label{2F_TOV2}
    \begin{aligned}
\frac{d p_{2}}{d r}=&-\frac{G M(r) \epsilon_{2}(r)}{r^{2}}\left[1+\frac{p_{2}(r)}{\epsilon_{2}(r)}\right] \left[1+4 \pi r^{3} \frac{\left(p_{1}(r)+p_{2}(r)\right)}{M(r)}\right]\left[1-2 G \frac{M(r)}{r}\right]^{-1} ,
\end{aligned}
\end{equation}
\begin{equation}
\frac{d M_{1}}{d r}=4 \pi r^{2} \epsilon_{1}(r)   , 
\end{equation}
\begin{equation}
    \frac{d M_{2}}{d r}=4 \pi r^{2} \epsilon_{2}(r) ,
\end{equation}
\begin{equation}\label{mass2_2F_TOV}
M(r)=M_{1}(r)+M_{2}(r).
\end{equation}
where $p_{i}$, $\epsilon_{i}$ and $M_{i}$ refer to the pressure, energy density and mass, respectively, and the indices $i=1,2$ refers to each of the fluids. To solve this system of differential equations we must give as inputs the equations of state for each of the fluids, the central densities (or pressures), $\epsilon_{1,c}$ and $\epsilon_{2,c}$, and values for the masses at the center, $M_{1}(0)=0$ and $M_{2}(0)=0$. 

Notice that the presence of poloidal and toroidal magnetic fields can modify the geometry of the stars \cite{Das2015_uso_das_TOV_1, Negreiros2018}, altering the structure equations discussed above. However, as argued in Refs. \cite{Das2015_uso_das_TOV_1, Deb2021}, the deviations from spherical symmetry are small, so that the assumption of spherical symmetry is a reasonable approximation.

\section{Equations of State}

\subsection{Magnetized strange quark matter}\label{sec: EoSs}

We consider strange quark matter (SQM) composed of massive quarks up, down and strange, besides electrons, and describe it within the MIT bag model framework. \footnote{There are more realistic models in the literature that usually allow for higher maximum masses e.g., Refs. \cite{Alford2005,Kurkela:2014vha}). The price to pay is usually the inclusion of additional parameters or bands of uncertainty. Those, although relevant and certainly more realistic, could obscure the effects on which we are focusing here: those that come from presence of a strong magnetic field and of dark matter. Since the MIT bag model is successful in capturing most of the qualitative behavior, we favor simplicity over a more realistic description,  postponing the latter for a future investigation.} We adopt a bag constant of $\mathcal{B}^{\frac{1}{4}}=145$ MeV, which yields a maximum mass of $M=2.01$ $M_{\odot}$ for strange stars. The main effect of the presence of a magnetic field is the appearance of Landau levels as discussed in Refs. \cite{chak1, chak2}.

Assuming a uniform magnetic field of the form $\vec{B}=B\hat{z}$, the energy density is given by 
\begin{multline}\label{eq:energy_matter}
    \epsilon_{m}=\frac{|B|}{4 \pi^{2}} \sum_{f} \sum_{\nu_{=0}}^{\nu_{\text {max }}} g_{f} q_{f}\Bigg[\mu_{f} \sqrt{\mu_{f}^{2}-M_{f, \nu}^2}
    +M_{f, \nu}^{2} \ln \left(\frac{\mu_{f}+\sqrt{\mu_{f}^{2}-M_{f, \nu}^2}}{M_{f, v}}\right)\Bigg]+\mathcal{B},
\end{multline}
and the pressure by
\begin{multline}\label{eq:pressure_matter}
    P_{m}=\frac{|B|}{4 \pi^{2}} \sum_{f} \sum_{\nu_{=0}}^{\nu_{\text {max }}} g_{f} q_{f}\Bigg[\mu_{f} \sqrt{\mu_{f}^{2}-M_{f, \nu}^2}-
    M_{f, \nu}^{2} \ln \left(\frac{\mu_{f}+\sqrt{\mu_{f}^{2}-M_{f, \nu}^2}}{M_{f, v}}\right)\Bigg]-\mathcal{B},
\end{multline}
where $f=u, d, s, e$, $q_f$ is the fermion electric charge, and $g_f$ is the degeneracy factor for the Landau levels ($1$ for the lowest and $2$ for the rest). The subscript $\nu$ corresponds to the Landau levels for each fermion and
\begin{equation}
    {M_{f, \nu}}\equiv\sqrt{m_{f}^2+2q_{f}|q_{f}B|\nu}.
\end{equation}
is an effective mass that depends on the intensity of the magnetic field. We obtain the total energy density and pressure by adding to the matter components the contributions brought about by the field, so that:
\begin{equation}\label{EoS_mag_1}
    \epsilon=\epsilon_{m}+\frac{B^2}{8\pi},
\end{equation}
\begin{equation}\label{EoS_mag_2}
    P=P_{m}-\frac{B^2}{8\pi}.
\end{equation}

One can see that the resulting equation of state is softer than the non-magnetic EoS for the bag model. Moreover, notice that we implicitly assumed that the pressure is isotropic, i.e., $P_{||}=P_{\perp}$. Since one can show that significant anisotropies appear only for fields of order $5\times 10^{18}$ G \cite{Huang2010} and we restrict our analysis to magnetic fields up to $2\times 10^{18}$ G, we take $P_{||}\approx P_{\perp}$ as a reasonable approximation. 

It is also worth mentioning that the relevant length scales involved in the microphysics of the EoS are very small (much smaller than the atomic radius), thereby insensitive to spatial variations of the magnetic field. Such variations can still be important for the gravitational setup but, as we discussed in section \ref{sec:structure_eq}, the distortions that they produce are small. Some authors also consider density-dependent magnetic fields, which means that the magnetic fields are stronger in the core of the star and weaker closer to the surface. To implement such magnetic fields, one must assume a given profile, even though there is not a general agreement on which magnetic field profile one should use (see Refs. \cite{Dexheimer2017} and \cite{Lopes2015} for examples of two different choices). Therefore, in this first attempt to describe strange magnetars admixed with dark matter, we stick to the simplest description: a constant magnetic field. This can be thought of as a limiting case, in which the highest value of the magnetic field extends to the whole star.

Except for the cases in which we compare results for different values of the magnetic field, a magnetic field of intensity $B=10^{18}$ G is assumed. This corresponds to an energy density of $B^2/8 \pi \simeq 24.8$ MeV fm$^{-3}$, which is of the same order of the bag constant $57$ MeV fm$^{-3}$. We  adopt a mass of $0.5$ MeV for the electrons, $5$ MeV for quarks up and down and $150$ MeV for strange quarks.

Since we are interested in discussing properties of compact stars, we must require chemical equilibrium and charge neutrality:
\begin{equation}
\begin{split}
    \mu_{d}=\mu_{s}=\mu, \\
    \mu_{u}+\mu_{e}=\mu,
\end{split}
\end{equation}
and
\begin{equation}
    \frac{2}{3}n_{u}-\frac{1}{3}n_{d}-\frac{1}{3}n_{s}-n_{e}=0, 
\end{equation}
where the densities have the form
\begin{equation}\label{full_fermion_density_mag}
     n_{f}=\frac{g_{f}|q_{f}B|}{2\pi^2}\sum^{\nu_{max}}_{\nu}\sqrt{\mu_{f}^2-M_{f, \nu}^2}
\end{equation}
for each fermion species. Notice that each sum over Landau levels may have different upper limits.

For comparisons with the non-magnetic case we use the massless quark limit of the MIT bag model, which corresponds to an equation of state of the form:
\begin{equation}
    p=\frac{1}{3}(\epsilon-4\mathcal{B}).
\end{equation}
%


\subsection{Fermionic dark matter}\label{sec_model_DM_2}

We consider two simple models for fermionic dark matter: a gas of free particles and a gas of strongly self-interacting particles \cite{Narain:2006kx,Mukhopadhyay:2015xhs}. For each case we consider dark fermion masses of $m=5,100,500$ GeV.
If we define $z\equiv k_{F} / m$, where $k_{F}$ is the Fermi momentum of the DM particles and $m$ its mass, we can write the expressions for the  energy density and pressure for a free Fermi gas of DM particles in a dimensionless form \cite{Narain:2006kx}:
\begin{equation}\label{energy_Fermi_gas_Ad}
\frac{\epsilon}{m^{4}} =\frac{1}{8 \pi^{2}}\left[\left(2 z^{3}+z\right) \sqrt{1+z^{2}}-\sinh ^{-1}(z)\right],
\end{equation}
\begin{equation}\label{press_Fermi_gas_Ad}
    \frac{p}{m^{4}} =\frac{1}{24 \pi^{2}}\left[\left(2 z^{3}-3 z\right) \sqrt{1+z^{2}}+3 \sinh ^{-1}(z)\right].
\end{equation}

For the self-interacting case, $\epsilon$ and $p$ are given simply by the sum of expressions (\ref{energy_Fermi_gas_Ad}) and (\ref{press_Fermi_gas_Ad}) with a term due to the self-interaction in the mean field approximation \cite{Narain:2006kx}. To make this contribution dimensionless, we define a coupling constant $\alpha_{\rm SI}\equiv g_{\rm v DM}^{2}/2$ and the interaction energy scale $m_{I}=m/\sqrt{\alpha_{\rm SI}}$. This energy scale, $m_{I}$, may also be interpreted as the vacuum expectation value of the Higgs field of the interaction, and the dimensionless parameter $y=m/m_{I}$ is a measure of the strength of the self-interaction. We can then write the energy density and pressure in terms of $z$ and $y$ as follows:
\begin{equation}\label{energy_Fermi_gas_Ad_SI}
\frac{\epsilon}{m^{4}} =\frac{1}{8 \pi^{2}}\left[\left(2 z^{3}+z\right) \sqrt{1+z^{2}}-\sinh ^{-1}(z)\right]
+\Big(\frac{1}{3\pi^2}\Big)^2 y^2 z^6,
\end{equation}
\begin{equation}\label{press_Fermi_gas_Ad_SI}
    \frac{p}{m^{4}} =\frac{1}{24 \pi^{2}}\left[\left(2 z^{3}-3 z\right) \sqrt{1+z^{2}}+3 \sinh ^{-1}(z)\right]
    +\Big(\frac{1}{3\pi^2}\Big)^2 y^2 z^6.
\end{equation}
 Following \cite{Mukhopadhyay:2015xhs}, we consider the case for which $y=0$ (free) and the case $y=10^3$ (strong self-interaction) with mass scale  comparable to the QCD energy scale. With the equations of state at hand, we can solve the TOV equation.

\section{Results for the mass and structure}\label{sec:results}

\subsection{Effect of magnetized strange quark matter on the dark matter component}

\begin{figure*}
    \centering
 \includegraphics[width=15cm]{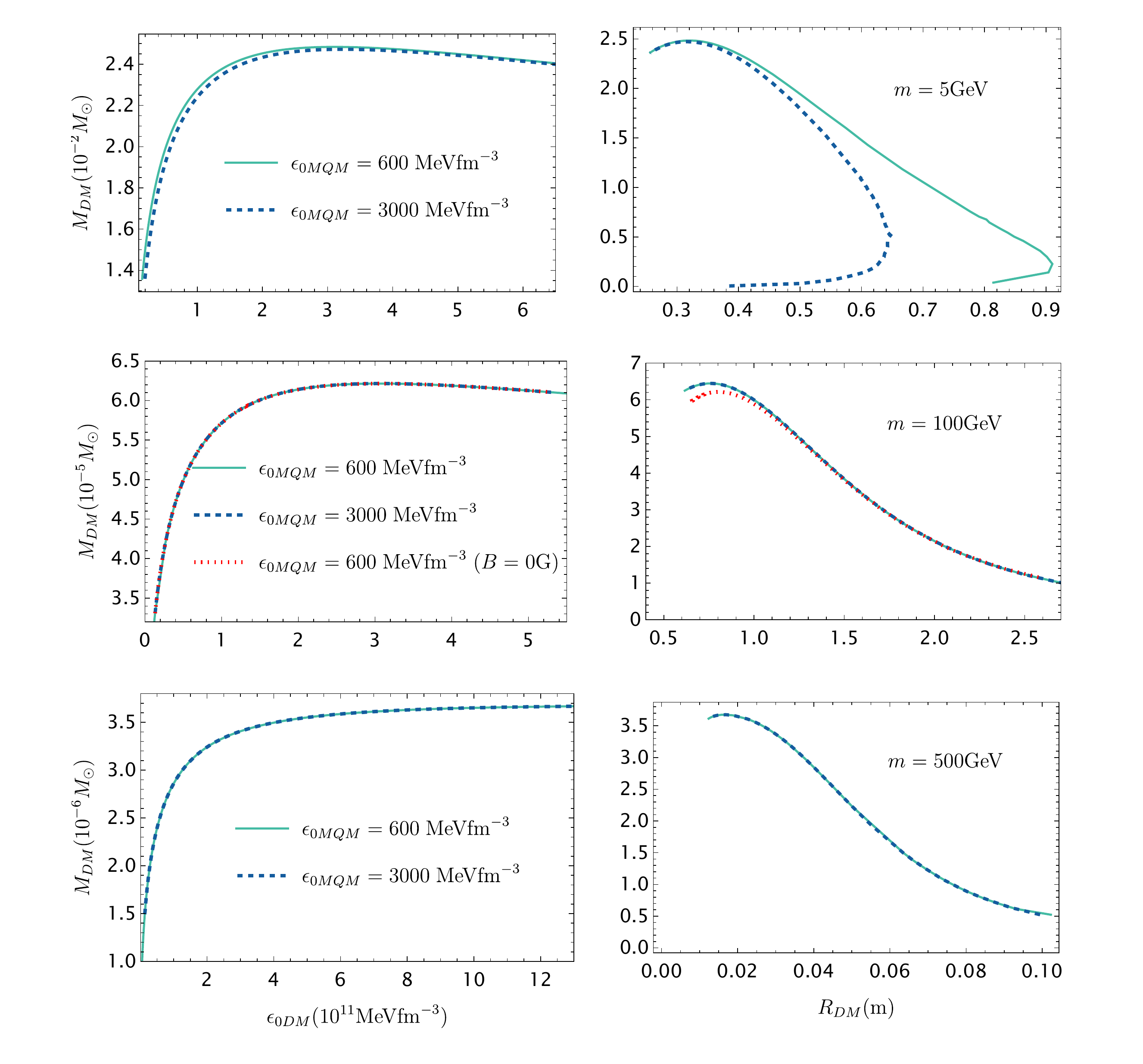}
    \caption{Effects of magnetized strange quark matter on the free dark matter component. In the first row we plot the mass of the dark matter component, $M_{DM}$, as a function of dark matter central energy density, $\epsilon_{0DM}$, for different fixed values of magnetized strange quark matter central energy density, $\epsilon_{MQM}$. In the second row we plot the mass-radius relations for different fixed values of $\epsilon_{MQM}$. We consider three values for the mass of the dark matter particles: $m=5,100,500$ GeV. The red dashed lines correspond no magnetic field ($B=0$ G). For all other curves, $B=10^{18}$G.}
    \label{panel_MDM_Free}
\end{figure*}

\begin{figure*}
    \centering
    \includegraphics[width=15cm]{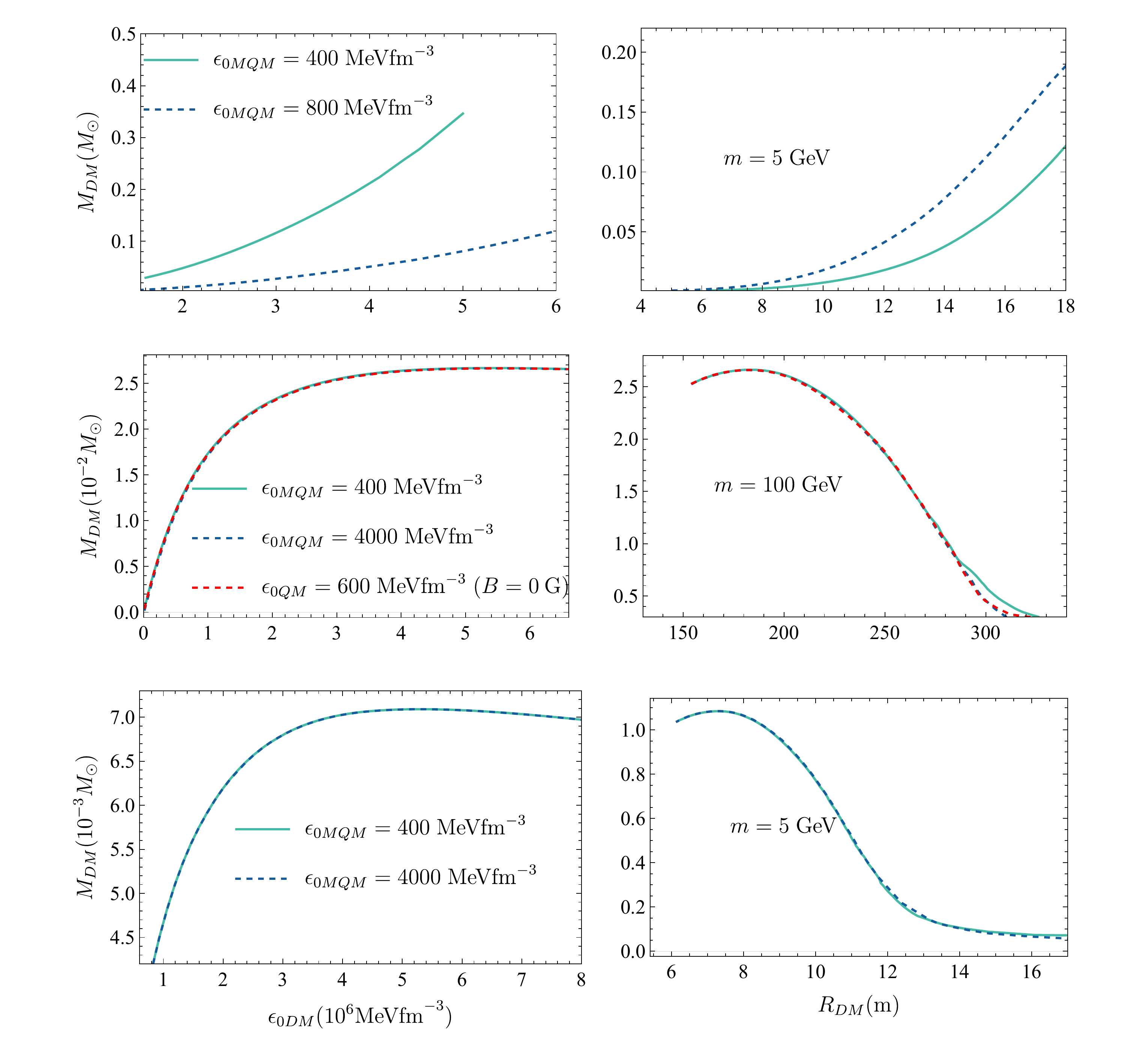}
     \caption{Effects of magnetized strange quark matter on the strongly self-interacting dark matter component. In the first row we plot the mass of the dark matter component, $M_{DM}$, as a function of dark matter central energy density, $\epsilon_{0DM}$, for fixed values of magnetized strange quark matter central energy density, $\epsilon_{0MQM}$. In the second row we plot the mass-radius relations for the dark matter component for fixed values of $\epsilon_{0MQM}$. We consider three values for the mass of the dark matter particles: $m=5,100,500$ GeV and assume a interaction strength of $y=10^3$. The red dashed lines correspond no magnetic field ($B=0$ G). For all other curves, $B=10^{18}$G.}
    \label{panel_MDM_INT}
\end{figure*}

In this section we discuss the effects of magnetized strange quark matter (MSQM) on the properties of the dark matter component of the star. The overall result is that MSQM hardly affects the dark matter component in both the free and strongly self-interacting (SSI) cases, except for a small dark fermion mass ($m=5$ GeV), as shown in Figs.\ref{panel_MDM_Free} and \ref{panel_MDM_INT}. This reinforces the results of references \cite{Mukhopadhyay:2015xhs} and \cite{Jimenez:2021nmr} for the non-magnetic case (red dashed lines in Figs.\ref{panel_MDM_Free} and \ref{panel_MDM_INT}). Essentially, the dark matter component is insensitive to the increase of SQM central energy density, when $m \gtrsim 100 $ GeV. Here we see that the same occurs for a softer magnetized EoS. 

Dark matter components made of particles with smaller masses ($m=5$ GeV) are more sensitive to the increase of MSQM central energy density. In the case of free DM, a larger $\epsilon_{0MQM}$ produces smaller radii for the DM component while the range of possible masses remains unchanged (see Fig.\ref{panel_MDM_Free}). In the SSI case, both mass and radius of the DM component are unbounded, with $M_{DM}$ vs. $\epsilon_{0DM}$ and mass-radius curves growing as we increase $\epsilon_{0MQM}$ (see Fig.\ref{panel_MDM_INT}). This unphysical behavior is probably due to unstable configurations at high values of $\epsilon_{0DM}$, as discussed for the non-magnetic case in Ref. \cite{Jimenez:2021nmr}.

We also observe from Figs.\ref{panel_MDM_Free} and \ref{panel_MDM_INT} that for smaller values of the mass of DM particles we obtain larger masses and radii for the DM component (in accordance with \cite{Jimenez:2021nmr, DCO_overview}). For instance, in the free DM case we find a maximum mass of $M_{DM}=0.024$ $M_{\odot}$, $M_{DM}=6.2\times 10^{-5}$ $M_{\odot}$ and $M_{DM}=3.6\times 10^{-6}$ $M_{\odot}$, for $m=5$ GeV, $m=100$ GeV and $m=500$ GeV, respectively. The presence of interaction also leads to more massive DM components. In the SSI DM case, we find for the maximum mass $M_{DM}=2.6\times 10^{-2}$ $M_{\odot}$ and $M_{DM}=7\times 10^{-3}$ $M_{\odot}$, for $m=100$ GeV and $m=500$ GeV, respectively.

\begin{figure*}
    \centering
\includegraphics[height=16cm]{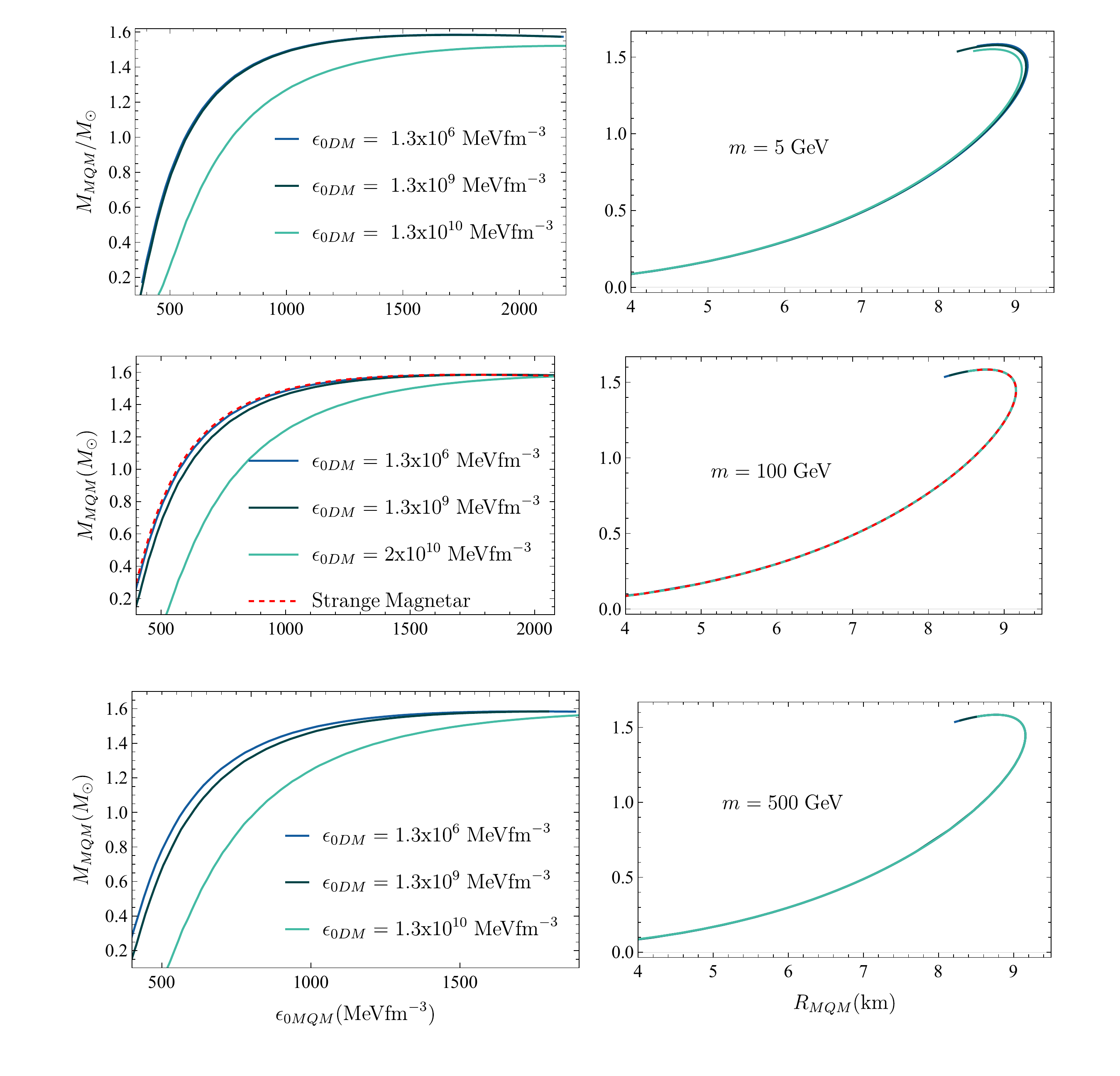}
    \caption{Effects of free fermionic dark matter on the properties of the magnetized strange quark matter component. In the first row we plot the mass of the magnetized strange quark matter component, $M_{MQM}$, as a function of magnetized strange quark matter central energy density, $\epsilon_{0MQM}$, for fixed values of dark matter central energy densities, $\epsilon_{0DM}$. In the second row we plot the mass-radius relations of the magnetized strange quark matter component for fixed values of $\epsilon_{0MQM}$. We consider three values for the mass of the dark matter particles: $m=5,100,500$ GeV and assume a interaction strength of $y=0$. The red dashed lines correspond to the strange magnetars with no dark matter.}
    \label{panel_MQM_free}
\end{figure*}

\begin{figure*}
    \centering
    \includegraphics[width=16cm]{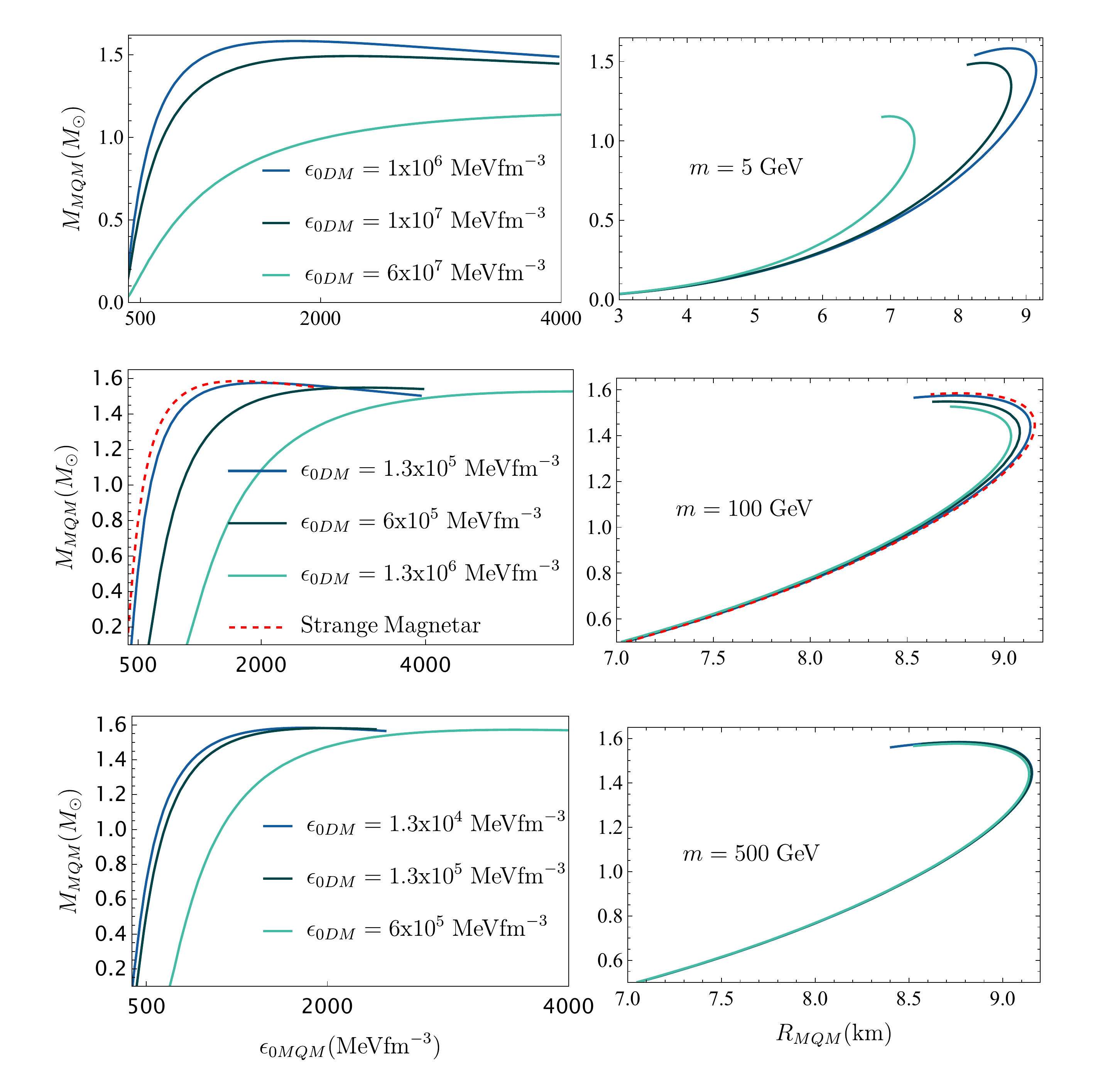}
    \caption{Effects of strongly self-interacting fermionic dark matter on the properties of the magnetized strange quark matter component. In the first row we plot the mass of the magnetized strange quark matter component, $M_{MQM}$, as a function of magnetized strange quark matter central energy density, $\epsilon_{0MQM}$, for fixed values of dark matter central energy densities, $\epsilon_{0DM}$. In the second row we plot the mass-radius relations of the magnetized strange quark matter component for fixed values of $\epsilon_{0MQM}$. We consider three values for the mass of the dark matter particles: $m=5,100,500$ GeV and assume a interaction strength of $y=10^3$. The red dashed lines correspond to the strange magnetars with no dark matter.}
    \label{panel_MQM_INT}
\end{figure*}




\subsection{Effect of dark matter on the magnetized strange quark matter component}

We now discuss the effects of dark matter on the magnetized strange quark matter component. A general feature is that, as we increase $\epsilon_{0DM}$, the curves of $M_{MQM}$(or $R_{MQM}$) versus $\epsilon_{0MQM}$ are displaced towards higher $\epsilon_{0MQM}$, and so are their maxima (see Figs. \ref{panel_MQM_free} and \ref{panel_MQM_INT}). In a first naive analysis of stability, the points before the maxima (from left to right) correspond to stable configurations. Therefore, this shifting feature implies that the range of stable configurations is also shifted. To illustrate this point, let us look at the $m=100$ GeV case. For strange magnetars (without dark matter) the maximum mass is reached for $\epsilon_{0MQM}=2700$ MeVfm$^{-3}$. When dark matter is added the maximum mass is reached when $\epsilon_{0MQM}=3940$ MeVfm$^{-3}$ for $\epsilon_{0DM}=1.3 \times 10^{5}$ MeVfm$^{-3}$ and when $\epsilon_{0MQM}=6000$ MeVfm$^{-3}$ for $\epsilon_{0DM}=1.3\times 10^{6}$ MeVfm$^{-3}$.

The maximum mass of the magnetized strange quark matter component is more sensitive to variations in central energy density of dark matter components made of lighter particles. This can be seen from Fig. \ref{panel_MQM_INT} which shows that the $m=5$ GeV case is the most affected, followed by the $m=100$ and $m=500$ GeV, respectively. In the free case, only for $m=5$ GeV one sees a slight decrease of the maximum mass (see Fig. \ref{panel_MQM_free}). We therefore conclude from Figs. \ref{panel_MQM_free} and \ref{panel_MQM_INT} that the MSQM component is more sensitive to variations in $\epsilon_{0DM}$ of SSI DM than of free DM. Finally, we should point out that the values of $\epsilon_{0DM}$ were chosen first aiming to illustrate the effects, and second to keep the configurations in the branch in which magnetized quark matter is dominant. In other words, since we are interested in magnetized strange quark stars we chose here to analyze the case in which dark matter is concentrated in the center of the quark star and not the opposite. As discussed in previous references, higher concentrations of dark matter leads to dark matter dominated compact objects \cite{Tolos:2015qra}.




\begin{figure*}
    \centering
 \includegraphics[height=12cm]{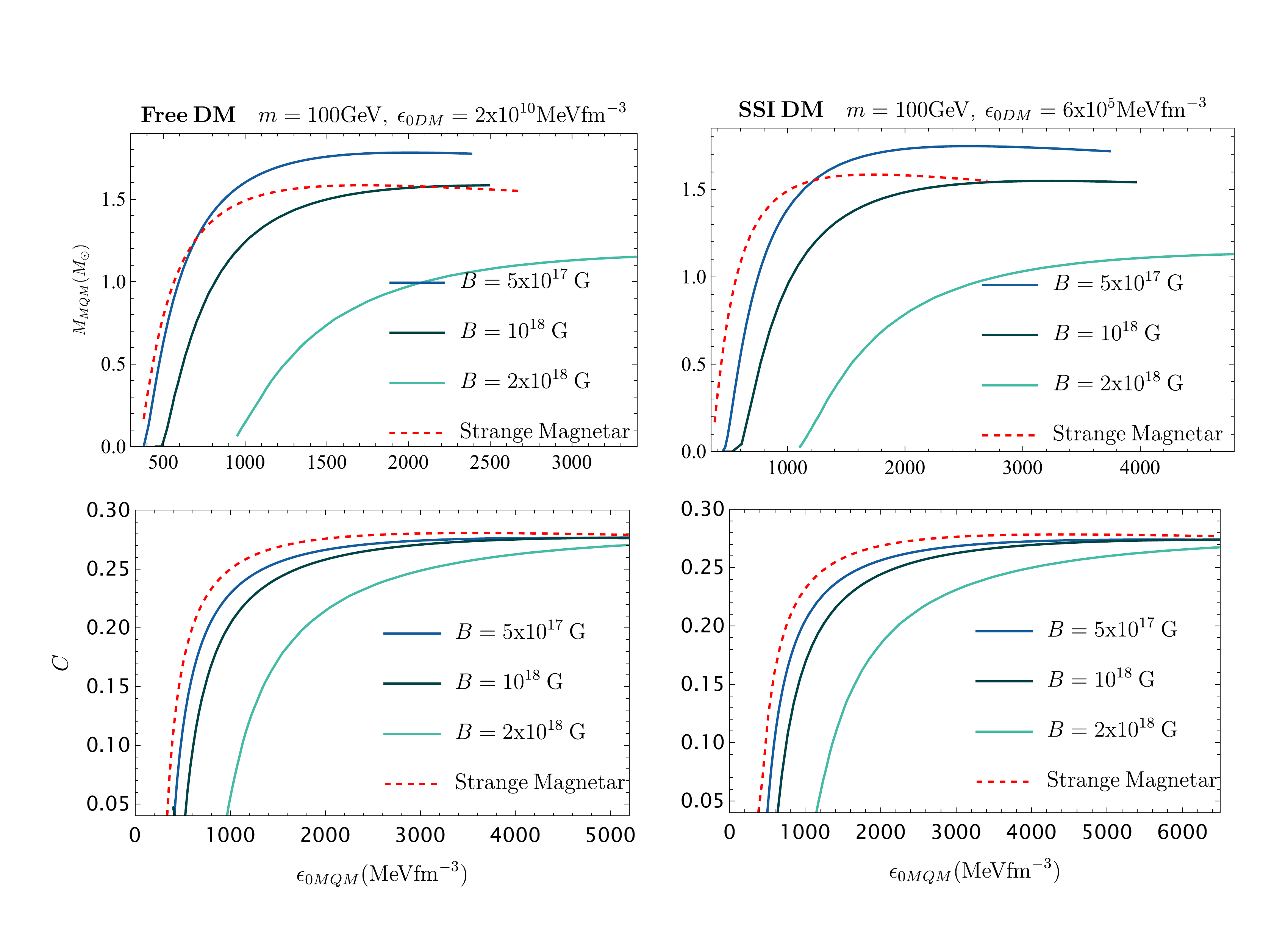}
    \caption{Mass and compactness ($C=G M_{T}/(c^2 R_{MQM})$) of the magnetized strange quark matter component as a function of its central energy density, $\epsilon_{0MQM}$, with fixed values of $\epsilon_{0DM}$, for different values of the magnetic field. Here $m=100$GeV. Left: free dark matter; right: strongly self-interacting dark matter. In the first row the red dashed line correspond to strange magnetars with no dark matter and in the second it corresponds to strange stars admixed with dark matter with no magnetic field.}
    \label{fig:Effect_mag_on_curves}
\end{figure*}

\subsection{Effects of dark matter on the maximum mass and the role of the magnetic field}

In previous work it was found that increasing the dark matter fraction decreases the total mass ($M\equiv M_{DM}+M_{MQM}$) of a system composed of SSI dark matter and SQM linearly \cite{Mukhopadhyay:2015xhs}. In this section we investigate if this still holds in the magnetized case and what is the role played by the intensity of the magnetic field. 

First, to investigate the effects of the field on the $M_{MQM}$ vs. $\epsilon_{0MQM}$ curves, we fix $m=100$ GeV and $\epsilon_{0DM}=2\times 10^{10}$ MeV fm$^{-3}$ for free DM and $\epsilon_{0DM}=6\times 10^{5}$ MeV fm$^{-3}$ for SSI DM. Notice from Figs. \ref{panel_MDM_Free} and \ref{panel_MDM_INT} that these values correspond to stable configurations of the DM component. 


\begin{figure}
    \centering
    \includegraphics[height=6cm]{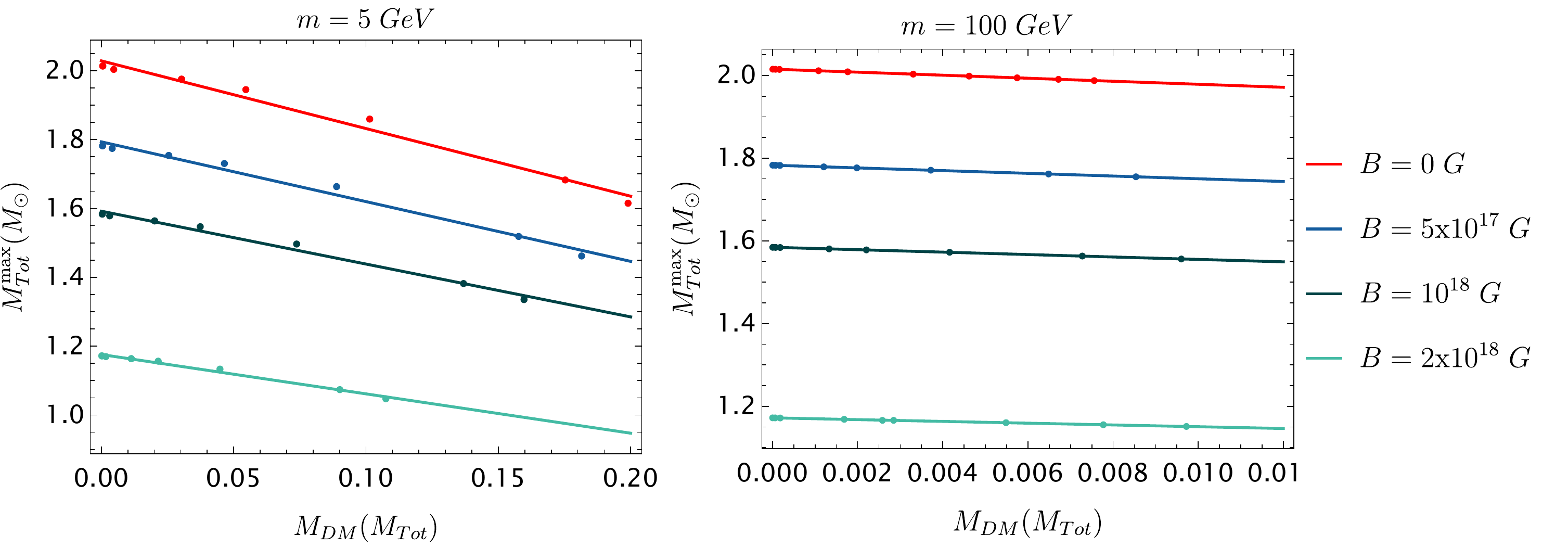}
    \caption{Maximum total mass of strange magnetars admixed with dark matter $M^{\mathrm{max}}_{T}$ as a function of dark matter fraction $M_{T}/M_{DM}$ for different values of magnetic field.}
    \label{fig:PlotMmaxvsDMfraction_diff_mag_all}
\end{figure}

From Fig. \ref{fig:Effect_mag_on_curves} we see that, as in the non-admixed case \cite{chak1}, increasing the magnetic field leads to smaller maximum masses. Something similar occurs for the radius, and the two effects balance each other yielding the same range of compactness 
\begin{equation}
C= \frac{G M_{T}}{c^2 R_{MQM}}
\end{equation}
for configurations with different magnetic fields. The non-magnetic case shows a slightly higher maximum compactness of $C_{\mathrm{max}}=0.278$ than the magnetized one, with $C_{\mathrm{max}}=0.274$. But no change is observed for the different magnetic fields of Fig. \ref{fig:Effect_mag_on_curves}. What changes is that stars with the same compactness support higher central energy densities when in the presence of a strong magnetic field. This is also observed in the curves $M_{MQM}$ vs. $\epsilon_{0MQM}$ which are shifted towards regions of higher $\epsilon_{0MQM}$. 

\vspace{0.2cm}

\begin{table}[t]
\begin{center}
    \begin{tabular}{|c|cl|}
        \hline
                                & \multicolumn{2}{c|}{Slope of the linear fit ($\alpha$)}\\
        \hline
        B (G)                  &  m=5 \text{GeV} & m=100 \text{GeV}\\
        \hline 
        $0$                     &   -1.9              & -3.6       \\
        $5 \times 10^{17}$ G    &    -1.7             & -3.2       \\
        $1 \times 10^{18}$ G    &     -1.5            & -2.9        \\
        $2 \times 10^{18}$ G    &     -1.1            & -2.1        \\
        \hline
    \end{tabular}
    \caption{Slopes of the linear decrease of the maximum mass with the dark matter fraction for different values of the magnetic field.}
    \label{tab:Slopes_Mtot_DMfraction}
\end{center}     
\end{table}

So, increasing the magnetic field produces a qualitatively similar effect to increasing the dark matter central energy density. The net result of including both a strong magnetic field and a dark matter component is a curve, $M_{MQM}$ vs. $\epsilon_{0MQM}$, that covers a range of higher $\epsilon_{0MQM}$ and has a reduced maximum mass. This curve is different from what is observed when only a magnetic field or a dark matter component is included. This is illustrated in Fig. \ref{fig:Effect_mag_on_curves} by comparing the curve for an admixed strange magnetar with the curve of a strange magnetar in the absence of dark matter (red dashed lines in Fig \ref{fig:Effect_mag_on_curves}) when the same value for the magnetic field is used ($B=10^{18}$ G). This difference occurs in both free and SSI DM. 

Finally, from Fig. \ref{fig:PlotMmaxvsDMfraction_diff_mag_all} we see that, as in the non-magnetized case, the total mass decreases linearly with the increase of the SSI dark matter fraction $f=M_{T}/M_{DM}$. A new feature that we observe here is that the rate of this decrease depends on the intensity of the magnetic field. As is shown in Table \ref{tab:Slopes_Mtot_DMfraction}, the absolute value of the slope of the linear fit is smaller for higher values of the magnetic field. So higher magnetic fields tends to attenuate the effect of increasing the dark matter fraction on the maximum mass of the system. 

Furthermore, comparing the $m=5$ GeV and the $m=100$ GeV cases, we see that the mass of the DM particle affects: (i) the range of DM fractions (lighter particles allow for higher DM fractions), and (ii) the magnitude of the slope of the curves. Clearly, then, {\it the resulting behaviour of the maximum mass as a function of DM fraction is a combination of effects from the mass of the DM particle and the intensity of the magnetic field}.

This dependence with the intensity of the magnetic field can be understood as follows. Magnetic fields shift stable configurations towards higher central energy densities. This high MSQM central energy densities make it harder for the dark matter core to exert its influence on the MSQM component properties. We therefore need larger variations in dark matter fraction to produce the same variations in total mass.

\section{Results for the tidal deformability}

To investigate more deeply the EoS of dense matter inside compact stars, one needs to go beyond the measurement of their masses and radii. An important window was open recently by the detection of gravitational waves (GW) from the merger of a binary neutron stars by the LIGO-Virgo Collaboration \cite{LIGO1, LIGO2}: the tidal deformation of the stars in the inspiral phase of the merger do affect the GW signal, bringing a new useful observable. 

Then, on one hand, the presence of dark matter in these binaries can modify their tidal deformability \cite{CS_DM_GWs1, CS_DM_GWs2, CS_DM_GWs3, CS_DM_GWs4, Tolos_k2}. On the other hand, magnetars are usually found isolated \cite{Review_Mag1, Review_Mag2}. However, since most massive stars are in binary systems, if core-collapse supernovae frequently give birth to magnetars, some fraction of them are expected to have a companion at the time of observation \cite{Magnetars_companions}. So, the merger of a system with at least one magnetar could provide us with useful information about their structure. 

Tidal deformability of strange quark stars have been discussed in a number of previous papers \cite{Albino:2021zml, Postnikov2010, Wang:2021jyo,Lopes2023arxiv} where the impact of interactions, the self-bound character of the EoS, and even the presence of dark matter were considered. Here we focus on the role played by both the magnetic field and dark matter on the resulting tidal deformability.

\subsection{Definitions}

In this section we investigate the behaviour of the tidal deformability of strange magnetars admixed with fermionic dark matter. The tidal deformability of a star characterizes the deformation of a star in response to the gravitational tidal field produced by its companion \cite{Hinderer1}. Here we briefly state the main equations (for details, see Refs. \cite{Hinderer1, Hinderer2, DamourBC}. 

We assume a static, spherical star placed in an external quadrupolar field $\mathcal{E}_{ij}$. To linear order, we have  
\begin{equation}
    \mathcal{Q}_{ij}=-\lambda \mathcal{E}_{ij},
\end{equation}
where $\mathcal{Q}_{ij}$ is the induced quadrupole moment and $\lambda$ is the tidal deformability \cite{Hinderer1, Hinderer2}. The tidal deformability $\lambda$ is related to the dimensionless second Love number
\begin{equation}
    k_{2}=\frac{3}{2}R^{-5},
\end{equation}
which can be calculated using 
\begin{equation}
\begin{aligned}
k_2= & \frac{8}{5} C^5(1-2 C)^2\left[2-y_R+2 C\left(y_R-1\right)\right] \left\{2 C\left(6-3 y_R+3 C\left(5 y_R-8\right)\right)\right. \\
& +4 C^3\left[13-11 y_R+C\left(3 y_R-2\right)+2 C^2\left(1+y_R\right)\right] \\
& \left.+3(1-2 C)^2\left[2-y_R+2 C\left(y_R-1\right)\right] \log (1-2 C)\right\}^{-1},
\end{aligned}
\end{equation}
where the function $y$ is obtained by solving the equation
\begin{equation}\label{eq_for_y}
    r y^{\prime}+y^2+y e^\lambda\left[1+4 \pi r^2(p-\rho)\right]+r^2 Q=0
\end{equation}
together with the TOV equations introduced in section \ref{sec:structure_eq}. $C$ is the compactness parameter and $y_{R}=y(r=R)$. The primes in equation (\ref{eq_for_y}) denote radial derivatives and the function $Q(r)$ is given by
\begin{equation}\label{Q_funtion}
    Q=4 \pi e^\lambda\left(5 \rho+9 p+\frac{\rho+p}{d p / d \rho}\right)-\frac{6 e^\lambda}{r^2}-\left[\frac{2\left(m+4 \pi r^3 p\right)}{r^2(1-2 m / r)}\right]^2.
\end{equation}
Moreover, we must use proper boundary conditions for self-bound stars as discussed in Refs. \cite{DamourBC, Hinderer2}. 

Finally, we define the dimensionless tidal deformability as
\begin{equation}
    \Lambda=\frac{2k_{2}}{3 C^5} \, .
\end{equation}

Since we are interested in the case of admixed stars, we must evaluate the tidal deformability of a system of two spherical fluids that interact only gravitationally. As discussed in \cite{CS_DM_GWs3, CS_DM_GWs4}, for the case of admixed stars we must make the replacement
\begin{equation}
    \frac{\rho+p}{d p / d \rho} \rightarrow \sum_i \frac{\rho_i+p_i}{d p_i / d \rho_i}
\end{equation}
on equation (\ref{Q_funtion}). The subscript $i$ refers to the dark and quark matter components. Notice that $d p/d\rho$ corresponds to the speed of sound squared, $c_{s}^2$. 
\begin{figure*}
    \centering
 \subfloat{\includegraphics[width=10cm]{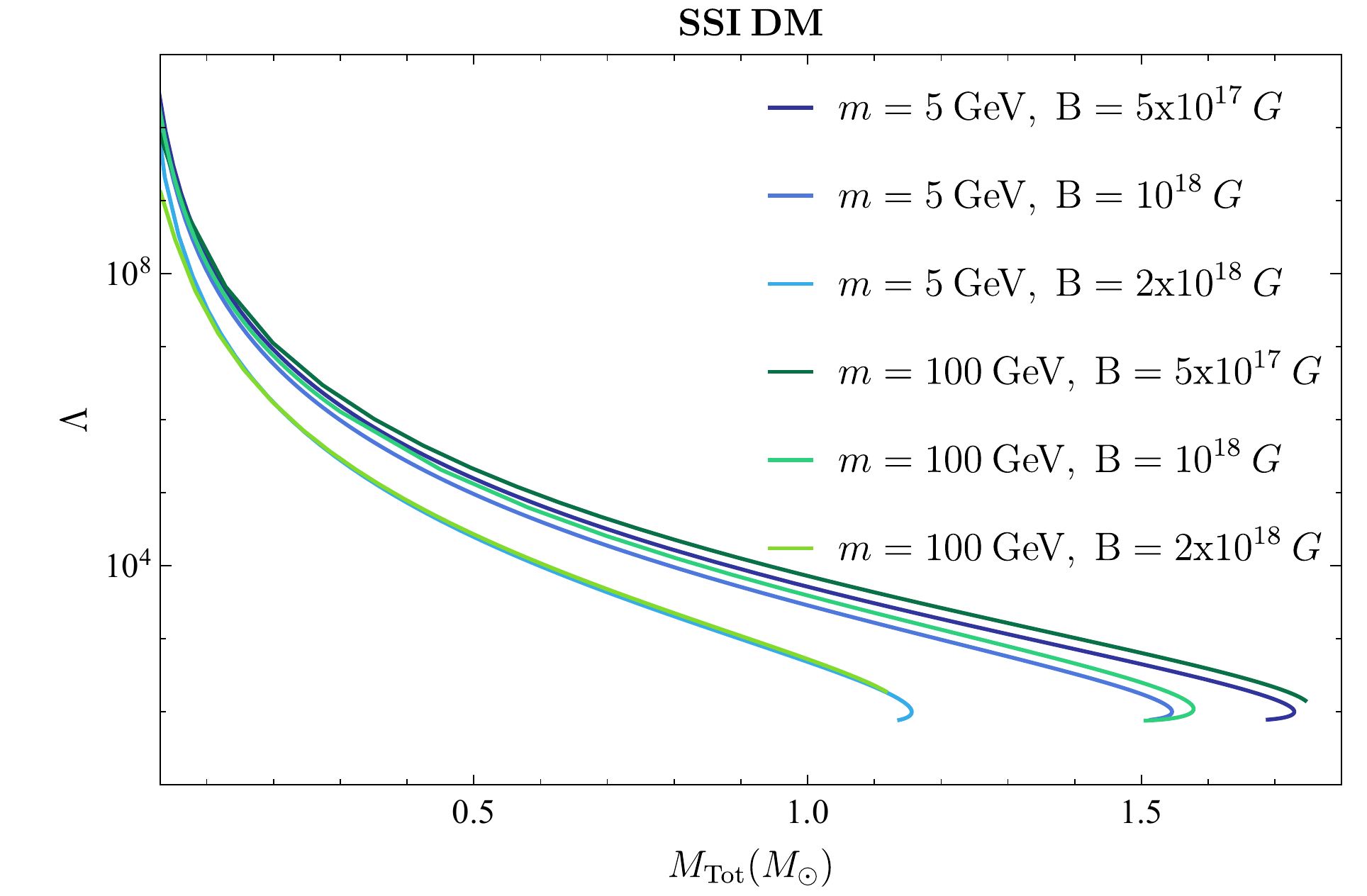}}
 \\
\subfloat{\includegraphics[width=10cm]{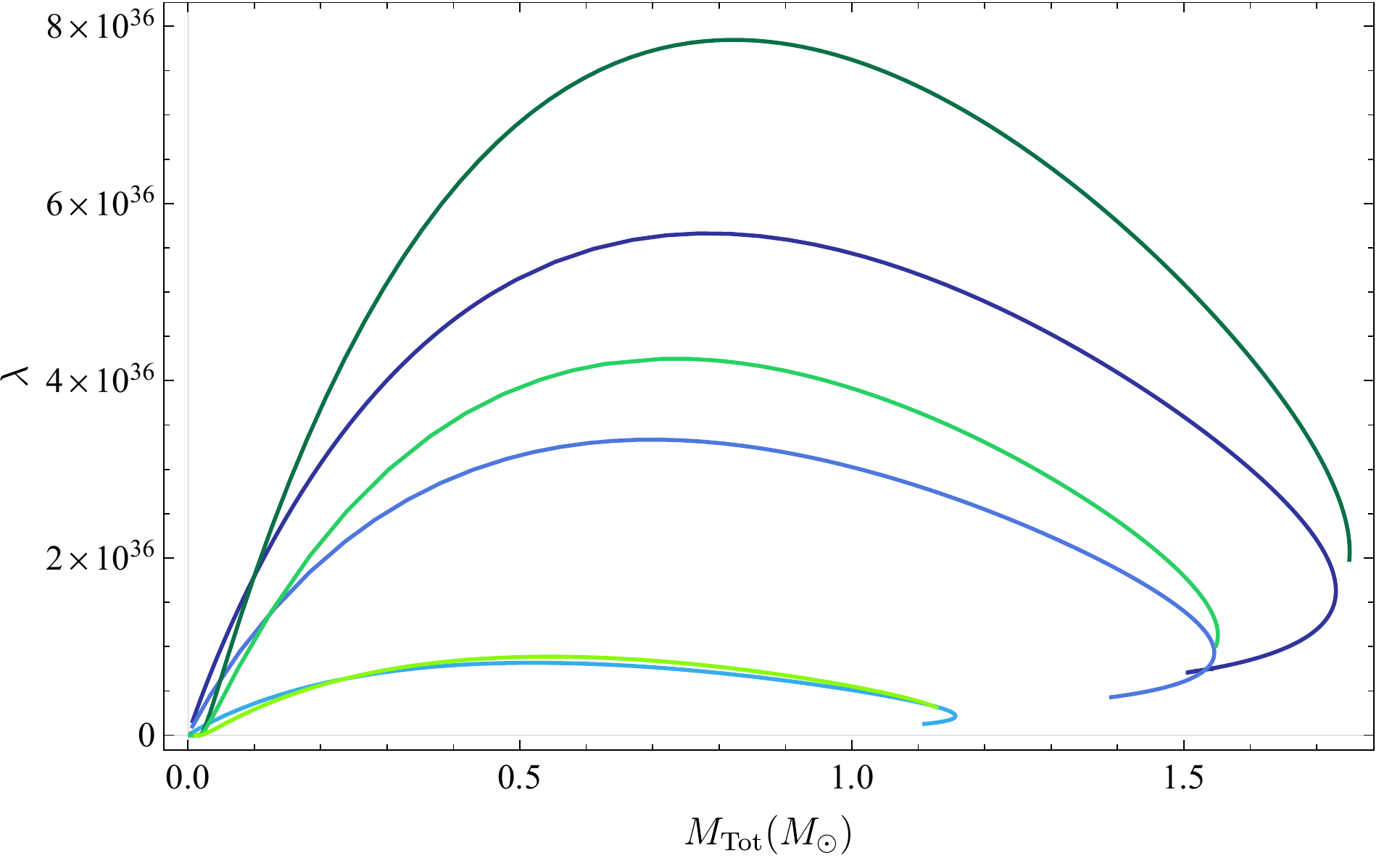}}
    \caption{Tidal deformability and deformability parameter for SSI DM for different values of magnetic field and fixed DM central energy density.  We fixed $\epsilon_{0DM}=1\times 10^{7}$ MeV fm$^{-3}$ for $m=5$ GeV  and $\epsilon_{0DM}=1\times 10^{6}$ MeV fm$^{-3}$ for $m=100$ GeV.}
    \label{fig:DeformabilityDMenergyDensityMag}
\end{figure*}



\begin{figure*}
    \centering
 \subfloat{\includegraphics[width=10cm]{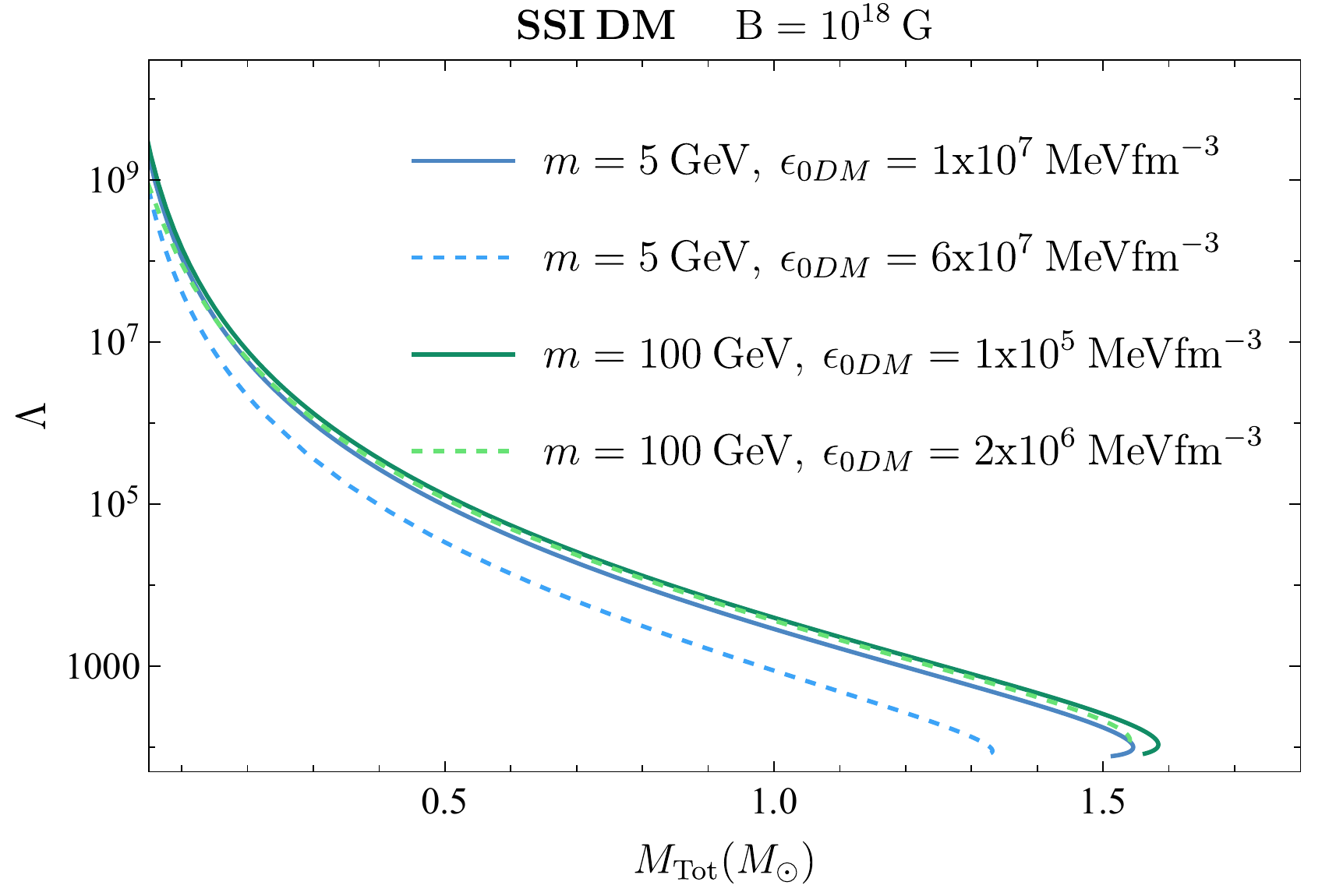}}
 \\
\subfloat{\includegraphics[width=10cm]{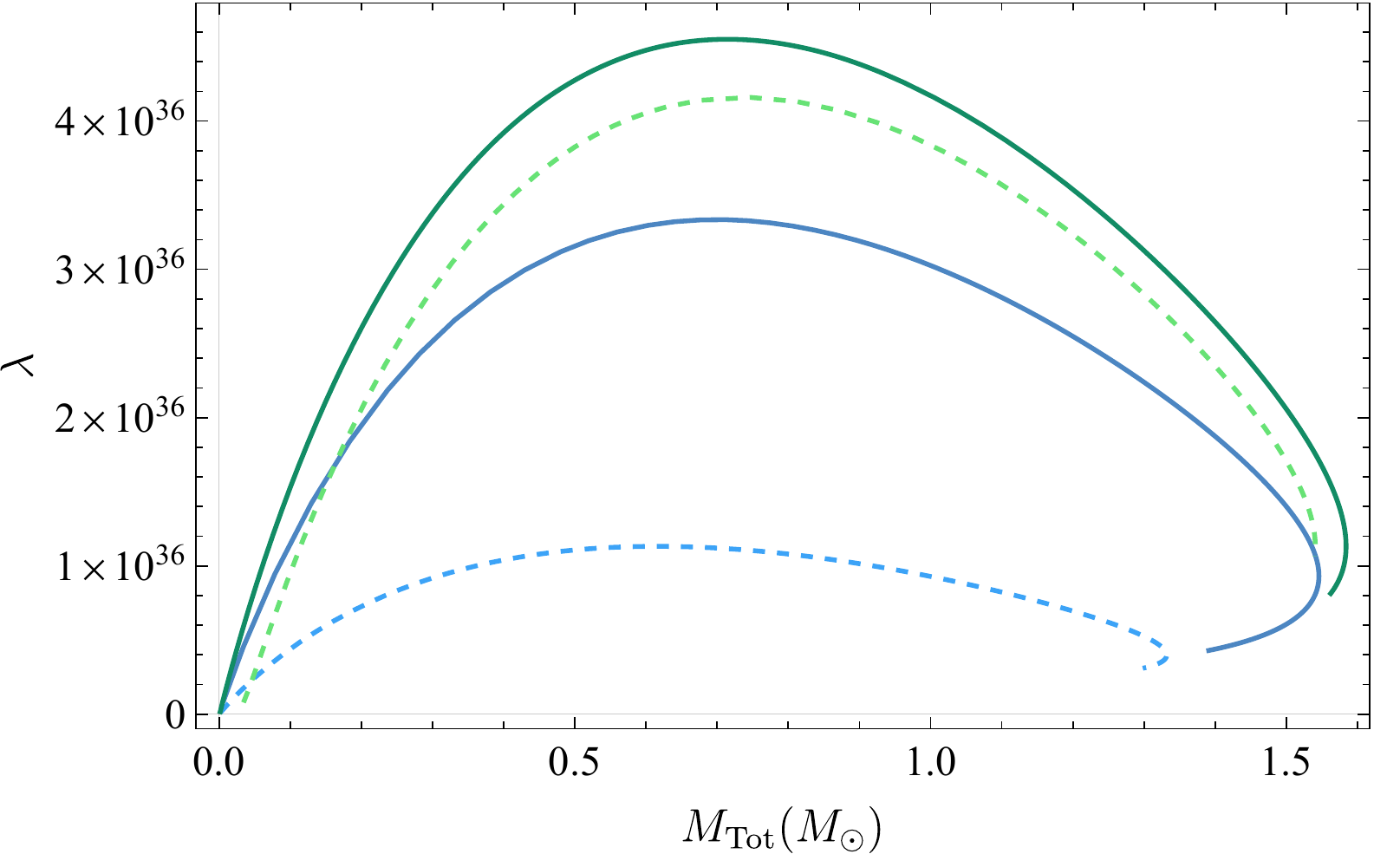}}
    \caption{Tidal deformability and deformability parameter for SSI DM for a fixed magnetic field of $B=10^{18}$ G.}
    \label{fig:DeformabilityDMenergyDensity}
\end{figure*}


\subsection{Results}

As could probably be anticipated by the results of the previous sections, the case of free DM, and even the case of SSI DM with $m=500$ GeV,  do not leave a relevant observable imprint on the tidal deformability and related quantities. These quantities are functions of compactness and thus of the total masses and radii of the stars. When these masses and radii are not appreciably modified, the parameters associated with the deformability are also essentially unaffected. For free DM with $m=5$ GeV there are small changes when DM central energy densities are increased, but these changes are still very small. We therefore focus on the cases of strongly self-interacting dark matter with $m=5$ GeV and $m=100$ GeV. 

Again, the previous sections give us a hint of what to expect: the presence of dark matter and of a strong magnetic field produce a similar effect. Thus, we can anticipate that they will also change the curves of the parameters associated with the deformability in a similar fashion. Looking at Figs. \ref{fig:DeformabilityDMenergyDensityMag} and  \ref{fig:DeformabilityDMenergyDensity} this is indeed what we observe. 

Increasing the intensity of the magnetic field leads to a decrease on both the dimensionless tidal deformability and the tidal deformability parameter (In Fig. \ref{fig:DeformabilityDMenergyDensityMag} we fixed $\epsilon_{0DM}=1\times 10^{7}$ MeVfm$^{-3}$ for $m=5$ GeV  and $\epsilon_{0DM}=1\times 10^{6}$ MeVfm$^{-3}$ for $m=100$ GeV). Moreover, since the magnetic field reduces the maximum mass, the curves approach zero faster for higher magnetic fields.

In Fig. \ref{fig:DeformabilityDMenergyDensity} we fixed $B=10^{18}$ G. As discussed previously, the effects of increasing dark matter central energy density are similar to what we observed when varying the magnetic field. Moreover, we see from Fig. \ref{fig:DeformabilityDMenergyDensity} that the effects are more pronounced for lighter particles, in accordance with the results from the previous sections. 

For the purpose of illustration, in Table \ref{tab:tidal_deformability_var_mag_field} we consider a star of $1 M_{\odot}$ and give the values for the tidal deformability for different values of magnetic field and central energy densities. One can see that going from zero magnetic field to $B=2\times 10^{18}$ G (which corresponds to a magnetic energy density of $99.4$ MeV fm$^{-3}$) can lead to a reduction of the tidal deformability of $\sim 91 \%$ and $\sim 95 \%$ for $m=5$ GeV and $m=100$ GeV, respectively. A less abrupt increase in the magnetic field intensity from $B=5\times10^{17}$ G to $B=2\times 10^{18}$ G (increase of about sixteen times in the magnetic energy density) can also generate a large decrease in the tidal deformability of $\sim 89 \%$ and $\sim 93 \%$ for $m=5$ GeV and $m=100$ GeV, respectively. 

When looking at the effects from dark matter, we see that increasing the DM central energy density in about twenty times for a mass $m=100$ GeV, produces a more modest reduction of the tidal deformability of $\sim 10 \%$. On the other hand, a variation of the DM central energy density of about six times for $m=5$ GeV leads to a large decrease of the tidal deformability of $\sim 70 \%$ for $m=5$ GeV.

\begin{table}[t]
\begin{center}
   \parbox{.45\linewidth}{
\centering    
    \begin{tabular}{|c|cl|}
        \hline
                                & \multicolumn{2}{c|}{$\Lambda$}\\
        \hline
        $B$ (G)                  &  $m=5$ GeV & $m=100$ GeV\\
        \hline 
        $0$                     &   5285              & 9633       \\
        $5 \times 10^{17}$ G    &    3976             & 6846       \\
        $1 \times 10^{18}$ G    &    2769            & 2734        \\
        $2 \times 10^{18}$ G    &     455           & 475       \\
        \hline
 \end{tabular}
\caption*{\textbf{(a)}}
}
\hfill
\parbox{.5\linewidth}{
     \begin{tabular}{|l|c|cl|}
        \hline
                                   & $\epsilon_{0DM}$ (MeV fm$^{-3}$$ )$     & \multicolumn{2}{c|}{$\Lambda$} \\ \hline 
        \multirow{2}{*}{$m=5$ GeV}        &       $1\times 10^{7}$          &        \multicolumn{2}{c|}{2769}                  \\ \cline{2-3} 
                                           &         $6\times 10^{7}$             &         \multicolumn{2}{c|}{775}             \\ \hline 
    \multirow{2}{*}{$m=100$ GeV}          &         $1\times 10^{5}$            &   \multicolumn{2}{c|}{3904}               \\  \cline{2-3} 
                                           &        $2\times 10^{6}$               &    \multicolumn{2}{c|}{3544}           \\  \hline
    \end{tabular}
\caption*{\textbf{(b)}}
} 
    \caption{Values of tidal deformability for a $M=1$ $M_{\odot}$ star. (a) Different values of magnetic fields; (b) dark matter central energy density. In (a) we fixed $\epsilon_{0DM}=1\times 10^{7}$ MeV fm$^{-3}$ for $m=5$ GeV  and $\epsilon_{0DM}=1\times 10^{6}$ MeV fm$^{-3}$ for $m=100$ GeV. In (b) we fixed $B=10^{18}$ G.}
    \label{tab:tidal_deformability_var_mag_field}
\end{center}     
\end{table}

\section{Summary and outlook}\label{sec:conclusion}

We have investigared strange stars admixed with fermionic dark matter in the presence of a strong magnetic field using the two-fluid Tolman-Oppenheimer-Volkov equations. Magnetized strange quark matter was described within the MIT bag model and we considered magnetic fields in the range $\sim 10^{17}-10^{18}$ G. For dark matter, we considered the cases of free and strongly self-interacting dark fermions with masses $m=5, 100, 500$ GeV. 

The effects of dark matter on strange magnetars are qualitatively similar to what was previously obtained for strange stars. We found that the increase in the dark matter central energy density produces a decrease in the maximum mass of the strange star, which resembles the effect of increasing the magnetic field. This similarity also occurs for the tidal deformability. Both dark matter and the magnetic field tend to reduce the tidal deformability. For instance, for a $1$ $M_{\odot}$ star we can have a reduction that can be as large as $95 \%$ and $70 \%$ from the effects of the magnetic field and of dark matter, respectively. In this sense, the presence of dark matter and that of a magnetic field bring effects that go in the same direction. So, one should be careful regarding the presence of strong magnetic field or dark matter, since one can be masqueraded by the other.

However, the net result of including both a dark matter component and a strong magnetic field is different from what is expected for strange stars or admixed strange stars, as illustrated by Fig. \ref{fig:Effect_mag_on_curves}. Furthermore the absolute value of the rate at which $M^{\mathrm{max}}_{T}$ decreases with the dark matter fraction $(M_{DM}/M_{T})$ is lower for higher values of the magnetic field (see Table \ref{tab:Slopes_Mtot_DMfraction} and Fig. \ref{fig:PlotMmaxvsDMfraction_diff_mag_all}). In other words, even though the magnetic field contributes to decreasing the total mass of the star, it attenuates the rate of decrease in the maximum mass brought about by the presence of dark matter.

\begin{acknowledgments}
We thank J.C. Jim\'enez for discussions. This work was partially supported by INCT-FNA (Process No. 464898/2014-5), CAPES (Finance Code 001), CNPq, and FAPERJ.
\end{acknowledgments}

\bibliographystyle{unsrtnat}
\bibliography{referencesJCAP}

\end{document}